\documentclass[sn-apa]{sn-jnl}

\usepackage{program}

\usepackage{xcolor} 
\usepackage{color, colortbl}
            
\definecolor{openblue}{cmyk}{1,0,0,0}

\definecolor{Gray}{gray}{0.9}

\usepackage[normalem]{ulem}

\jyear{2021}%

\theoremstyle{thmstyleone}%
\theoremstyle{thmstyletwo}%

\theoremstyle{thmstylethree}%

\begin{document}

\title[Article Title]{Detecting Code Injections in Noisy Environments Through EM Signal Analysis and SVD Denoising}


\author[1]{\fnm{Ekaterina} \sur{Miller}}\email{ekat2834@vandals.uidaho.edu}

\author*[2]{\fnm{Georgios Michail} \sur{Makrakis}}\email{gmakrakis@uidaho.edu}

\author[2]{\fnm{Kurt A.} \sur{Vedros}}\email{kvedros@uidaho.edu}

\author[2]{\fnm{Constantinos} \sur{Kolias}}\email{kolias@uidaho.edu}

\author[3]{\fnm{Craig} \sur{Rieger}}\email{craig.rieger@inl.gov}

\author[4]{\fnm{Daniel} \sur{Barbara}}\email{dbarbara@gmu.edu}

\affil[1]{\orgdiv{Department of Computer Science}, \orgname{University of Idaho}, \orgaddress{\street{1031 N Academic Way}, \city{Coeur d'Alene}, \postcode{83814}, \state{ID}, \country{USA}}}

\affil*[2]{\orgdiv{Department of Computer Science}, \orgname{University of Idaho}, \orgaddress{\street{1776 Science Center Dr}, \city{Idaho Falls}, \postcode{83402}, \state{ID}, \country{USA}}}

\affil[3]{\orgdiv{National and Homeland Security}, \orgname{Idaho National Lab}, \orgaddress{\street{1955 N Fremont Ave}, \city{Idaho Falls}, \postcode{83402}, \state{ID}, \country{USA}}}

\affil[4]{\orgdiv{Department of Computer Science}, \orgname{George Mason University}, \orgaddress{\street{4400 University Dr}, \city{Fairfax}, \postcode{22030}, \state{VA}, \country{USA}}}


\abstract{The penetration of embedded devices in networks that support critical applications has rendered them a lucrative target for attackers and evildoers. However, traditional protection mechanisms may not be supported due to the memory and computational limitations of these systems. Recently, the analysis of electromagnetic (EM) emanations has gathered the interest of the research community. Thus, analogous protection systems have emerged as a viable solution e.g., for providing external, non-intrusive control-flow attestation for resource-constrained devices. Unfortunately, the majority of current work fails to account for the implications of real-life factors, predominantly the impact of environmental noise. In this work, we introduce a framework that integrates singular value decomposition (SVD) along with outlier detection for discovering malicious modifications of embedded software even under variable conditions of noise.  
Our proposed framework achieves 
high detection accuracy i.e., above 93\% AUC score for unknown attacks, even for extreme noise conditions i.e., -10 SNR. To the best of our knowledge, this is the first time this realistic limiting factor, i.e., environmental noise, is successfully addressed in the context of EM-based anomaly detection for embedded devices.}




\maketitle
\section{Introduction}\label{introduction}

Today, embedded devices have become an indispensable component of a wide range of heterogeneous applications.
Such devices are deployed to support even mission-critical tasks of Industrial Control Systems (ICS) within Critical Infrastructures (CI). Therefore, it does not come as a surprise that embedded devices 
have become targets of cyberattacks. Nonetheless, due to their limited on-chip processing capabilities and reliance in proprietary OS and toolchains, the installation of traditional means of protection such as anti-malware or Host-based Intrusion Detection Systems (H-IDS) is deemed impractical. 
Because of this, the applicability of external means of control-flow attestation has attracted the interest of the research community. While several alternative solutions have been considered, the monitoring of systems based on \emph{side-channel analysis} has gained traction. 
Among the different types of side-channels (power \cite{liu2016code}
acoustic \cite{anton2019discussing}, thermal \cite{islam2017exploiting}), the analysis of electromagnetic (EM) signals is preferable because it offers high bandwidth and enables the monitoring of the Central Processing Unit (CPU) activity at fast sampling rates \cite{nazari2017eddie},\cite{han2017watch}.

Previous work in the area \cite{vedros2021limits}
\cite{nazari2017eddie},\cite{han2017watch}
has shown that it is possible to achieve high accuracy for detecting \emph{significant} violations in the execution flow of a monitored program under \emph{controlled laboratory environments}. However, minimal code injections of just a few instructions may be more stealthy and challenging to detect especially in noisy environments. 

In this work, we propose a framework for reliably identifying code injection attacks of \emph{arbitrary} lengths of code even under the influence of \emph{variable} levels of environmental noise. At the core of the framework lies a non-destructive, noise-reduction technique that is based on the Singular Value Decomposition (SVD) method. 
The analysis process is not based on supervised Machine Learning (ML), but rather it is based on an outlier detection strategy at the core of which lies the well-known Local Outlier Factor (LOF) \cite{breunig2000lof} method. This design decision renders the system capable of detecting even \emph{unseen/unknown attacks} i.e.
the injection of variable lengths of code and alternative types of instructions, that have not been analyzed before.

\section{Technical Background \& Definitions}
\label{sec:technical-background-definitions}

The execution of instructions by the CPU of a given device results in changes in the flow of electric current inside the CPU's circuitry. This change produces a magnetic field that interacts with the electric field, resulting in an EM field. Moreover, the components of the printed circuit board act as antennas. Thus, the board unintentionally transmits EM signals that are highly correlated to the instructions running at the CPU. These emanations can be captured by placing a probe near the source of the signal \cite{han2017watch}. 
It is generally accepted that it is possible to identify (at bare minimum) the active \emph{execution state} of a program by analyzing such analog signals. 

\begin{figure}[h]
  \centering
  \includegraphics[width=0.6\linewidth]{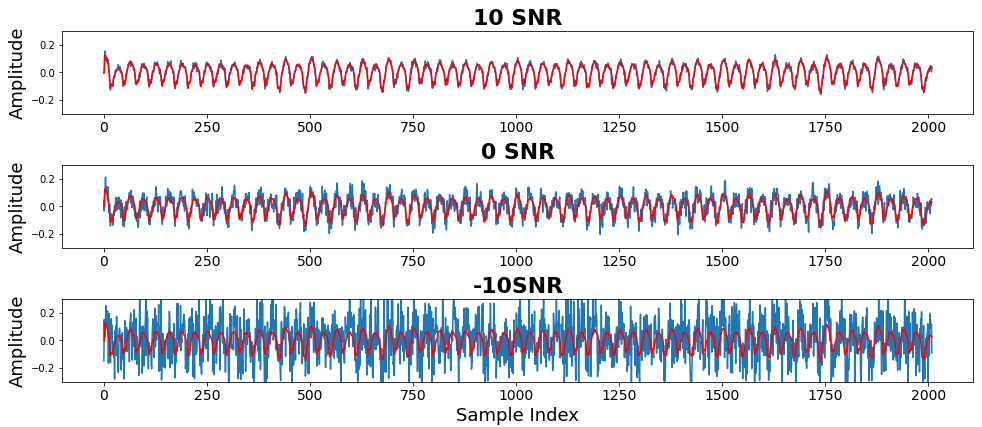}
  \caption{Comparison between a clean signal (red) and the noisy version of the same signal (blue) for various SNR levels.} 
  \label{fig:signals_diffrentSNR2}
\end{figure}

Code injection attacks start by exploring a target software for pre-existing vulnerabilities that allow an attacker to inject malicious code. During the exploitation step, the instructions that are injected change the original logic of the program or forge a new (malicious) execution path, thus altering the original control flow. Attacks based on \emph{buffer overflows} are examples of such malicious activity.

EM signals obtained by the CPU are amplitude modulated according to corresponding instruction, with a carrier signal that has a base frequency, that the clock of the monitored CPU \cite{khan2018detailed}. Therefore, different instructions can theoretically be distinguished by observing the amplitude of the signal across time. 
While previous research in the field has shown that is possible to detect as little as one-instruction injection with an average AUC rate of 98.7\% for low noise environments  \cite{vedros2021limits}, a decrease in the detection rate should be expected when the noise level increases. Figure \ref{fig:signals_diffrentSNR2} displays the same signal at different noise levels. It is clear that important signal artifacts get concealed behind noise for the low SNR conditions. At the same time, the utilization of noise filtering might affect the anomaly detection process due to the destructive nature of corresponding methods. 

\begin{figure}[h]
  \centering
  \includegraphics[width=0.6\linewidth]{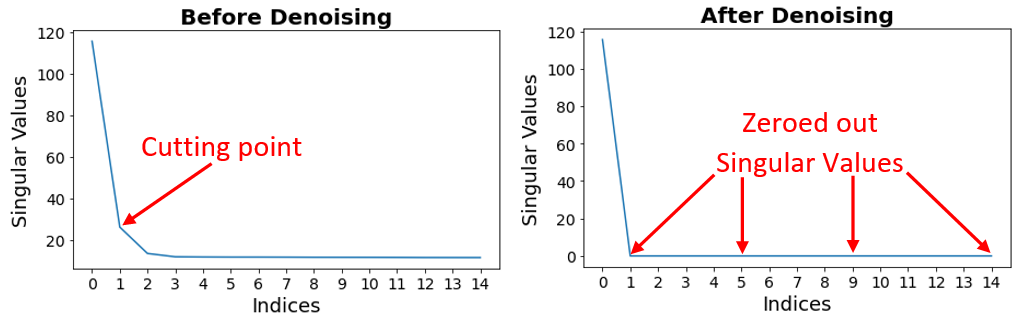}
  \caption{Position of the cutting point where the slope of the curve of the corresponding singular values changes drastically. Noise reduction can be achieved by zeroing out singular values after that point.}
  \label{fig:cutting_point}
\end{figure}

In this respect, previous work in the area \cite{govindarajan2019partial}, 
\cite{vedros2021limits} shows that SVD denoising is superior to other noise elimination techniques. Briefly, SVD decomposes a 2-D matrix into three components (matrices $U$, $\Sigma$, $V$). $\Sigma$ is a diagonal matrix where the values along the main diagonal correspond to the singular values of the 2-D matrix. In the context of noisy signals we assume that the 2-D matrix is the composition of a clean signal and noise. Then each of the $U$, $\Sigma$, $V$ can still be expressed as the composition of two signal subspaces:  the clean signal and the noise. In order to denoise the signal, all singular values corresponding to noise must be set to zero, thus creating a new approximation of the singular matrix, $\Sigma_{new}$. Figure \ref{fig:cutting_point} (right) illustrates this transformation. After this, a denoised version of the signal $X_s$ can be obtained as the dot product of $U$, $\Sigma_{new}$, and $V^T$. More specifically, this can be expressed as in Equation \ref{eq:basic-denoised}.

\begin{equation}
\begin{split}
\label{eq:basic-denoised}
X_s & = U \Sigma_{new} V^T = U \begin{bmatrix} \sigma_s & 0 \\ 0 & \sigma_n = 0\end{bmatrix} V^T 
\end{split}
\end{equation}

\noindent
where $X_s$ is the clean signal, $U$ is the left singular matrix, $\sigma_s$ are the singular values that correspond to the signal, $\sigma_n$ are the singular values that correspond to noise and $V^T$ is the right singular matrix transposed. One challenge with this approach is to accurately pinpoint the number of singular values corresponding to the signal. In other words, identifying the point in the $\Sigma$ matrix that accurately partitions the signal and the noise subspaces. That point is referred to as \emph{Cutting Point}. Hassanpour et al. \cite{hassanpour2012time} identified this point as the one where the slope of the curve of the corresponding singular values changes drastically.

\section{Proposed Framework}
\label{sec:proposed-framework}

Our proposed framework specifies practices for: (a) obtaining the EM signals from the subject device, (b) performing signal pre-processing to reduce noise while maintaining important characteristics that correspond to anomalies, (c) fingerprinting the morphology of EM signals corresponding to normal operations (training), and finally (d) quantifying the level of disparity of these signals versus the ones obtained on the field towards recognizing anomalous operations. 


\noindent
\textbf{Signal Acquisition:} To obtain the signals, the antenna must be placed in close proximity to the target device. The distance depends mainly on the size of the CPU chip and the enclosure/chassis of the device.
The collection of a large number of samples that correspond to each potential execution path of instructions (e.g., same loop) is needed to achieve efficient detection. From empirical data, the number of signal-observations of the same operation, should be at least the same as the number of samples in each signal-observation during the fingerprinting phase. During the deployment phase, this number should be double.
According to Vedros et al. \cite{vedros2021limits} high sampling rates, i.e., around x16 the CPU clock speed (with higher sampling rates providing diminishing returns), are ideal for anomaly detection purposes, while at the same time they provide robustness against noise and can reliably detect even small code injections.


\noindent
\textbf{Noise Reduction \& Pre-processing:} In this work, we apply a modification of the SVD denoising as a preprocessing step to reduce the noise level in the captured signals. The applied adaptations revolve around two axes. On the one hand, the method is applied to \emph{multiple examples} (as opposed to a single) describing the same phenomenon, i.e., the same execution path. This significantly speeds up the denoising process and alleviates the need for bringing individual observations to their Hankel Matrix representation as done in \cite{hassanpour2012time}, thus reserving memory. On the other hand, the signals are denoised by considering a \emph{higher-than-usual values for cutting points}. Although this decision may appear counter-intuitive, the reader should keep in mind that anomaly detection and not noise elimination is the main objective.
Therefore, since the only characteristic that distinguishes different instructions is the difference in the amplitude of the signal, we must take special care in preserving those differences during the denoising process.


\begin{algorithm}
{\footnotesize
\begin{algorithmic}[1]
\caption{Anomaly Detection Algorithm}
\label{alg:stroud}
\Function{Detect}{benign dataset $X$, test observation $q$}
\State $status_q \gets 0$

\For{$\forall i \in X$}
\State $s_{x_i} \gets LOF(X, x_i)$
\State $S_i \gets s_{x_i}$
\EndFor

\State $s_q \gets LOF(X, q)$

\For{$\forall s_{xi} \in S$}
    \If{$s_{x_i} \geq s_q$}
        \State $n \gets n + 1$
    \EndIf
\EndFor

\State $p_q \gets \frac{n+1}{|x|+1}$

\State $\tau \gets 1 - confidence$
\If {$p_{max} <= \tau$}
\State $status_q \gets 1$
\EndIf

\Return{$status_q$}

\EndFunction
\end{algorithmic}
}
\end{algorithm}

\noindent
\textbf{Anomaly Detection:} Our EM signal anomaly detection strategy is a modification of the transduction and hypothesis testing algorithm introduced by Barbara et al. in \cite{Barbara06}. 
The basic assumption of this method is that benign instances are easy to obtain in real-life conditions. Meanwhile, malicious cases, while bearing differences compared to benign observations, are highly unpredictable regarding the exact location, the morphology, and the extent of the disparity. Thus, the anomalous cases that can be observed in real-life may potentially be infinite. The latter is the primary reason why we chose to approach the problem at hand as an outlier detection problem rather than a classification problem. 
Transduction is carried out by placing an unknown signal in a known sample distribution of data and then carrying out hypothesis testing; it determines whether that instance is a good fit.
The process described above is given in Algorithm \ref{alg:stroud}. In further detail, the proposed approach requires that the following steps are executed:

\noindent
\emph{Step 1:} Collect a population of normal signals. Here, we assume that during the fingerprinting step all examples correspond to a benign operation and the device is not infected. This will be the \emph{baseline} of signals. 

\noindent
\emph{Step 2:} Apply the LOF algorithm for each example in the baseline to quantify the strangeness (i.e., the level of unfitness) of that signal with the rest. The reader may recall that LOF has been proven to have superior discrimination power over simpler methods like K-Nearest Neighbors (KNN) due to the fact that it takes into consideration the density of the neighboring points relative to the density of the point in consideration that is why it became the method of choice. This will be the \emph{baseline strangeness distribution.}

\noindent
\emph{Step 3:} During the deployment phase, obtain a signal we wish to test, and compute its \emph{unfitness score} with respect to the baseline constructed in Step 1 by using LOF. Then, transduce that value against the baseline strangeness distribution that was constructed in the previous step.

\noindent
\emph{Step 4:} As a result of the previous step, one obtains a fraction of the number of signals whose unfitness score is greater or equal to the unfitness score of the test point to the total signals considered. This fraction can be considered a \emph{p-value} for a statistical hypothesis test, where the \emph{null hypothesis} is this point belongs to the baseline distribution (i.e., it can be considered normal), while the alternative hypothesis is this point does not fit the distribution, (i.e. it needs to be deemed an anomaly).


\begin{figure}[h]
  \centering
  \includegraphics[width=0.4\linewidth]{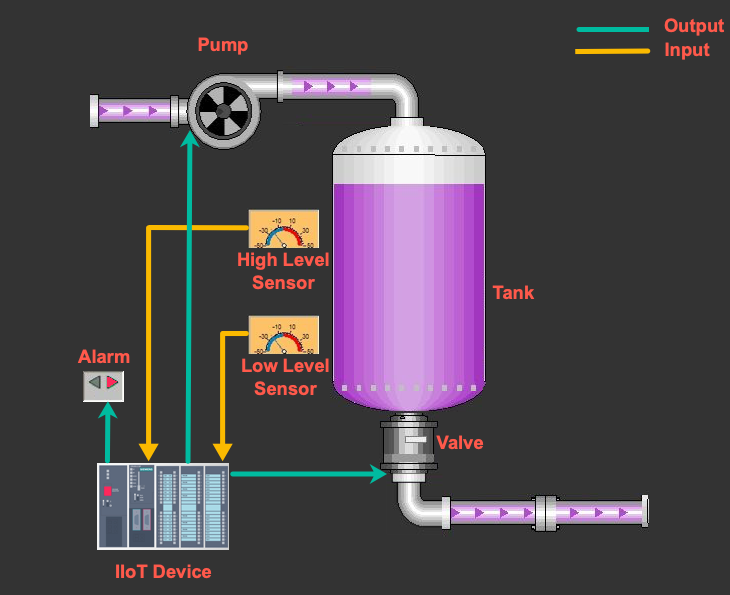}
  \caption{A graphical representation of described tank filling scenario.}
  \label{fig:controlled_process_2}
\end{figure}

\section{Experimental Setup}
\label{sec:experimental-setup-and-data-gathering}


To evaluate the proposed framework, we created an experimental setup consisting exclusively of low-cost, off-the-shelf components. A simple control process emulating a \emph{tank filling system} was used as the software-to-be-monitored for malicious modifications (Figure \ref{fig:controlled_process_2}). The control logic was implemented in the AVR assembly language and installed in an Arduino Mega with an ATmega2560 CPU clocked at 16MHz. The choice of language was made to have better control over the actual instructions being executed at the CPU.

The two adversarial cases considered in this work include the injection of ADD and JMP instructions, respectively. An ADD instruction consumes one CPU cycle, while a JMP instruction takes three cycles. Naturally, the injection of these malicious instructions causes a displacement of one and three cycles.

\begin{figure}[h]
    \centering
         \includegraphics[width=0.7\linewidth]{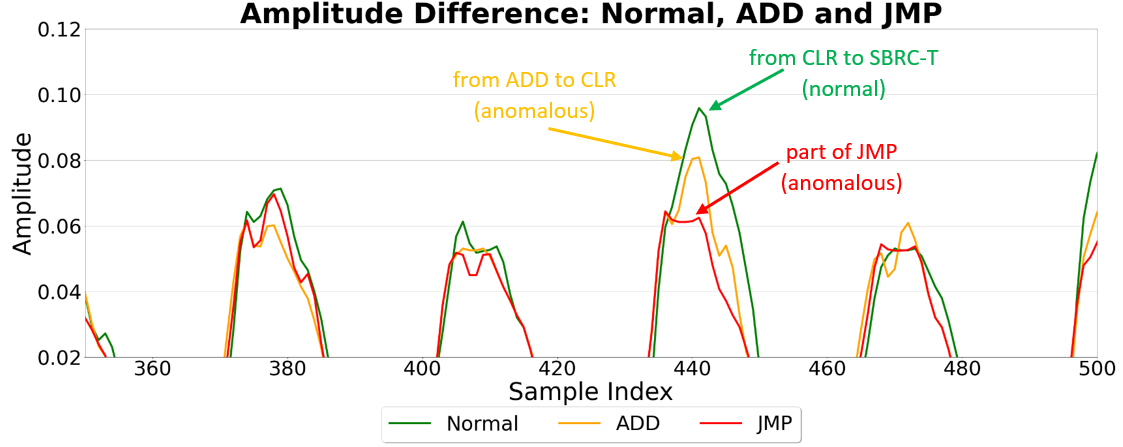}
  \caption{Zoom-in at the critical section where the injection occurred. The reader should compare the difference amplitude for each instruction that was injected vs the normal instruction.
  }
  \label{figure:normal-anomalous-injection-comp}
\end{figure}

\noindent
\textbf{Observations:}
By comparing the amplitude of the signals we can verify that there is a difference between the normal (expected CLR) and malicious instructions (ADD and JMP) (Figure \ref{figure:normal-anomalous-injection-comp}). 

\noindent
\textbf{Data Gathering:} We used a near-field probe placed directly on top of the device's CPU. This was done to obtain EM readings that are virtually noise-free. Later on, synthetic random noise having a Gaussian distribution was added to each signal. We considered noise levels of 10 SNR, 5 SNR, 0 SNR, -5 SNR, and -10 SNR. 

\noindent
\textbf{Evaluation:} To evaluate the detection accuracy of the proposed system, we used the ten-fold cross-validation method. More specifically, 90\% of the normal dataset was withheld and used for training, and 10\% of the remaining normal dataset, along with the same amount of anomalous signals, was used for testing purposes in each of the folds.
The average from the folds was calculated and reported. The predictive accuracy was evaluated using the rea Under Curve (AUC) score of the 
corresponding receiver operator characteristic (ROC) curve. 


\section{Experimental Evaluation}
\label{sec:experimental-evaluation}

Four sets of experiments were conducted to evaluate the efficiency of our framework.
The reader should keep in mind that all experiments consider the injection of a minimum number of instructions, i.e., one. Therefore, all results presented in this section define a lower bound (minimum) of predictive accuracy.

 \subsection{Considering No Pre-processing for Noise Reduction}

\begin{table}[h]
\caption{Performance of the detection algorithm without applying noise reduction as pre-processing. In parentheses, the number of neighbors provided as input to the anomaly detection algorithm.
} 
\label{tab:result_diffrent_no_SVD}
\begin{center}
\begin{tabular}{ |p{0.6cm}||p{1.7cm}|p{1.7cm}|p{1.7cm}|p{1.7cm}|p{1.7cm}|p{1.7cm}| }
 \hline
 \multicolumn{6}{|c|}{\textbf{Injection of ADD}} \\
 \hline
\textbf{SNR} & \textbf{10} & \textbf{5} & \textbf{0} & \textbf{-5} & \textbf{-10}\\
 \hline
    & 99.23\%(3) & 98.74\%(3) &	61.93\%(3) & 31.72\%(3) &
	21.90\%(3) \\

 \hline
  \multicolumn{6}{|c|}{\textbf{Injection of JMP}} \\
 \hline
 \textbf{SNR} & \textbf{10} & \textbf{5} & \textbf{0} & \textbf{-5} & \textbf{-10}\\
 \hline
 

     & 99.40\%(3) &	99.10\%(3) &	39.36\%(3) & 15.30\%(3) & 9.97\%(3)
  \\
  \hline
 
\end{tabular}
\end{center}

\end{table}

 As a first experiment, we wanted to identify the performance of the anomaly detection algorithm when no noise elimination is applied. 
Table \ref{tab:result_diffrent_no_SVD} contains the results (AUC scores) of the experiments. 
The results indicate that the AUC score is near-perfect when the system is deployed in a relatively clean environment (i.e., 10 to 5 SNR). 
However, the predictive accuracy rapidly drops to very poor levels (below 70\%) when considering SNR below 5dB. 

 \noindent
\textbf{Conclusions}: \emph{A noise reduction step is necessary, especially when the system is expected to operate at SNR levels below 5 dB.}

 \subsection{Considering SVD with Traditional Cutting Points for Noise Reduction}

  \begin{table}[h]
\caption{Performance of the detection algorithm (AUC score) considering SVD with traditional choice of cutting points, as a noise reduction step.}
\label{tab:result_diffrent_SVD_cp=1}
\begin{center}
\begin{tabular}{ |p{0.6cm}||p{1.7cm}|p{1.7cm}|p{1.7cm}|p{1.7cm}|p{1.7cm}|p{1.7cm}| }
 \hline
 \multicolumn{6}{|c|}{\textbf{Injection of ADD}} \\
 \hline
 \textbf{SNR}&  \textbf{10} & \textbf{5} & \textbf{0} & \textbf{-5} & \textbf{-10}\\
  \hline
   &  4.49\% (1,1,3) &	5.06\% (1,1,3)  &	5.49\% (1,1,3)   & 4.91\% (1,1,3) 	& 5.38\% (1,1,3)\\

 \hline
 \multicolumn{6}{|c|}{\textbf{Injection of JMP}} \\
 \hline
\textbf{SNR}&  \textbf{10} & \textbf{5} & \textbf{0} & \textbf{-5} & \textbf{-10}\\
 \hline
    &   4.46\% (1,1,3) &	4.98\% (1,1,3)  &	5.49\% (1,1,3)   & 4.87\% (1,1,3) 	& 5.60\% (1,1,3)\\

 \hline

\end{tabular}
\end{center}

\end{table}
 
As a next experiment, we introduced an additional pre-processing step aiming at noise reduction. For this experiment we assumed that the noise levels remain the same between fingerprinting and deployment phases. 
In this experiment, the parameter \emph{cutting point} is chosen according to traditional criteria i.e., where the slope of the curve in the corresponding singular values graph changes drastically \cite{hassanpour2012time}. 
Notice that based on the chosen criteria in all of our experiments the value of the cutting point was dictated to be $1$ (i.e., only on singular value is retained).

The results of this experiment are given in Table \ref{tab:result_diffrent_SVD_cp=1}. In parentheses the value of parameters (a) cutting point used in training phase, (b) cutting point used in deployment phase, and (c) number of neighbors. Based on the results, the application of SVD with traditional cutting points drastically degrades the anomaly detection accuracy for all cases. 
Apparently, such an aggressive noise reduction procedure eliminates important characteristics that are indicative of the anomalies.  

\noindent
\textbf{Conclusion}: \emph{The application of SVD for noise reduction as a pre-processing step when using traditional cutting points, has a negative impact on the anomaly detection process. Therefore, new criteria for choosing optimal cutting points should be identified.}

\subsection{Statistically Inferring Cutting Points for Anomaly Detection}

\begin{table}[h]
\caption{Performance of the detection algorithm (AUC score) considering SVD with alternative cutting points, as a noise reduction step.}

\label{tab:result_same_noise}
\begin{center}
\begin{tabular}{ |p{0.6cm}||p{1.7cm}|p{1.7cm}|p{1.7cm}|p{1.7cm}|p{1.7cm}|p{1.7cm}| }
 \hline
 \multicolumn{6}{|c|}{\textbf{Injection of ADD}} \\
 \hline
\textbf{SNR} & \textbf{10} & \textbf{5} & \textbf{0} & \textbf{-5} & \textbf{-10}\\
 \hline
    & 99.54\% (25,25,3) & 99.46\% (15,15,3) &	98.84\% (10,10,3) & 97.45\% (6,6,5) &
	92.20\% (4,4,5) \\

 \hline
  \multicolumn{6}{|c|}{\textbf{Injection of JMP}} \\
 \hline
 \textbf{SNR} & \textbf{10} & \textbf{5} & \textbf{0} & \textbf{-5} & \textbf{-10}\\
 \hline
 

     & 99.84\% (25,25,3) &	99.51\% (15,15,3) &	99.00\% (10,10,3) & 97.47\% (6,6,5) & 93.07\% (4,4,5)
  \\
  \hline
 
\end{tabular}
\end{center}

\end{table}

In the next set of experiments, we use a brute force approach to find the optimal cutting points across all considered noise environments. By increasing the value of the cutting point parameter, we are able to achieve a significant increase in predictive accuracy. Nevertheless, a dichotomy exists. \emph{Discarding} a lot of singular values may lead to the elimination of important characteristics indicative to anomalies, while \emph{retaining} too many may result in retaining noise that may conceal or distort these artifacts-of-interest. Detailed results for each experiment, along with the chosen hyperparameters, are given in Table \ref{tab:result_same_noise}.

\begin{figure}[h]
  \centering
  \includegraphics[width=0.5\linewidth]{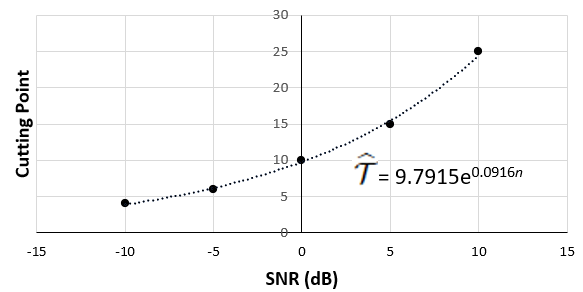}
  \caption{Statistical analyses of cutting points based on the noise level.}
  \label{fig:cp_equation}
\end{figure}

An interesting observation is that the values for the cutting points and the number of neighbors are the same, regardless of the types of injections. This indicates that the identified parameters for a given type of injection at a specific level of noise are potentially \emph{transferable} across different types of injections. By plotting those values (Figure \ref{fig:cp_equation}), we can statistically derive an Equation \ref{eq:cp_equation} that estimates the value of the optimal cutting point based on the active noise level. More specifically, the formula is given as:

 \begin{equation}
\widehat{\mathcal{T}} = 9.7915 \cdot e^{0.0916  \cdot  n}
 \label{eq:cp_equation}
 \end{equation}

\noindent
where 
$n$ is the noise level in the training and deployment environments; $\widehat{\mathcal{T}}$ is the value for the cutting points that need to be applied to reduce noise in the training and deployment environments. This is a number of singular values that will be retained during SVD denoising.

\noindent
\textbf{Conclusions:} \emph{(a) SVD denoising with cutting points that lead to the retention of more singular values provides high predictive accuracy even for minimal code injections and for extremely high levels of environmental noise; (b) It is possible to statistically infer a formula for identifying (near-optimal) cutting points for anomaly detection, regardless of the type of injection; (c) The cutting points identified through the formula achieved an appropriate level of denoising regardless on the type of injection.}


 

 


\subsection{Evaluating Considering Variable Noise Levels}
\label{sec:variable_Noise_levels}

 \begin{table}[h]
\caption{AUC scores across different noise levels.}


\label{tab:result_diffrent_noise_predict}

\begin{center}
\begin{tabular}{|p{0.4cm}|p{0.5cm}|p{1.2cm}|p{1.2cm}|p{1.2cm}|p{1.2cm}|p{1.2cm}| }
 \hline
 \multicolumn{7}{|c|}{\textbf{Injection of ADD}} \\
 \hline
\multicolumn{2}{|c|}{\textbf{test SNR}} &\textbf{10} & \textbf{5} & \textbf{0} & \textbf{-5} & \textbf{-10}\\
 \hline
 \rowcolor{Gray}
  \cellcolor{white} & \textbf{10} & \cellcolor{openblue!15}99.54\% (25,25,3) &	98.91\% (25,15,3) &	97.92\% (25,10,3) & 95.71\% (25,6,3)	& 86.25\% (25,4,25)\\
 \cellcolor{white} & \textbf{5} & - &	\cellcolor{openblue!15}99.46\% (15,15,3) & 98.02\% (15,10,3) &	96.67\% (15,6,5) &	86.88\% (15,4,25)\\
\rowcolor{Gray}
 \cellcolor{white}& \textbf{0} & - &	- & \cellcolor{openblue!15}98.84\% (10,10,5) & 96.89\% (10,6,5) & 87.17\% (10,4,25)\\
\cellcolor{white} &\textbf{-5} & - &	- & - & \cellcolor{openblue!15}97.45\% (6,6,5) & 87.28\% (6,4,51)\\
 \rowcolor{Gray}
    \cellcolor{white}\multirow{-8}*{\rotatebox{90}{\textbf{training SNR}}}& \textbf{-10} & - &	- & - & - & \cellcolor{openblue!15}92.20\% (4,4,5)\\

 \hline
 \multicolumn{7}{|c|}{\textbf{Injection of JMP}} \\
 \hline
\multicolumn{2}{|c|}{\textbf{test SNR}} &\textbf{10} & \textbf{5} & \textbf{0} & \textbf{-5} & \textbf{-10}\\
 \hline
 \rowcolor{Gray}
  \cellcolor{white} & \textbf{10} & \cellcolor{openblue!15}99.84\% (25,25,3) &	98.97\% (25,15,3) &	98.12\% (25,10,3) & 96.20\% (25,6,3)	& 88.37\% (25,4,25)\\
 \cellcolor{white} & \textbf{5} & - &	\cellcolor{openblue!15}99.51\% (15,15,3) & 98.33\% (15,10,3) &	96.94\% (15,6,5) &	88.72\% (15,4,25)\\
\rowcolor{Gray}
 \cellcolor{white}& \textbf{0} & - &	- & \cellcolor{openblue!15}99.00\% (10,10,3) & 96.99\% (10,6,5) & 88.90\% (10,4,25)\\
\cellcolor{white} &\textbf{-5} & - &	- & - & \cellcolor{openblue!15}97.47\% (6,6,5) & 88.94\% (6,4,51)\\
 \rowcolor{Gray}
    \cellcolor{white}\multirow{-8}*{\rotatebox{90}{\textbf{training SNR}}}& \textbf{-10} & - &	- & - & - & \cellcolor{openblue!15}93.07\% (4,4,5)\\
\hline

\end{tabular}
\end{center}

\end{table}

The subsequent set of experiments aims to quantify the predictive accuracy of our approach assuming that the noise levels have increased from the fingerprinting to the testing phase. The results of this experiment are provided in Table \ref{tab:result_diffrent_noise_predict}. In parentheses are the hyperapameters used to obtain the results namely, the cutting point during the fingerpinting phase, the cutting point during the deployment phase, and lastly the number of neighbors. The reader may notice that different cutting points were used for the two phases. All cutting points were derived automatically from Equation \ref{eq:cp_equation} without further analysis. Utilizing the new strategy for inferring cutting points dramatically increases the anomaly detection accuracy to levels higher than 95\% (AUC score) for the majority of the cases. 
The only exception to this rule is the case of having an extremely noisy deployment environment i.e., -10SNR. In this case, the best condition is to train the baseline in an equally noisy environment which yields an AUC score of 92.20\% and 93.07\% respectively. For the cases where the noise levels differ, an average of 86.89\% for the ADD and 88.73\% for the JMP is expected.
These results constitute a dramatic improvement over applying SVD noise reduction with conventional cutting points. 

\noindent
\textbf{Conclusions:} \emph{The proposed framework can achieve highly accurate anomaly detection even if the noise levels drastically increase from fingerprinting to deployment stage.}

\section{Conclusion}
\label{sec:conclusions-future-work}

EM-based anomaly detection systems may be particularly advantageous in the realm of embedded devices because they are able to detect modifications in software such as the injection of foreign instructions, remotely, without burdening the target system. However, the majority of works in the area, do not consider the impact of environmental noise. 
In this work, we introduced an EM-based anomaly detection framework that is robust against environmental noise. Through experiments, we proved that the proposed system yields highly accurate predictions i.e., ~94-95\% AUC score even in environments of extremely high noise levels, by utilizing the SVD technique to achieve noise reduction. 
Among the most important contributions of this work is an equation that can infer values for the highly sensitive denoising hyperparameters (i.e., cutting points).


\bmhead{Acknowledgments}

Prepared as part of the Laboratory Directed Research and Development program 
under DOE Idaho Operations Office Contract DE-AC07-05ID14517. 
We also acknowledge the Resilient Control and Instrumentation Systems (ReCIS) program. 

\bibliography{sn-article}


\end{document}